%% file: FBK3D_arXiv_V120319.tex
\documentclass[final,5p,times,twocolumn]{elsarticle}

\usepackage{graphicx}
\usepackage{amssymb}
\usepackage{subfigure}

\journal{Nuclear Instruments and Methods A}

\begin{document}

\begin{frontmatter}

\title{Characterization of proton irradiated 3D-DDTC pixel sensor prototypes fabricated at FBK}

%%%%%%%%%%%%%%%%%%%%%%%%%%%%%%%%%%%%%%%%%%%%%%%%%%%%%%
\author[A]{A. La Rosa\corref{cor1}}
\ead{alessandro.larosa@cern.ch}
\author[B]{M. Boscardin}
\author[C]{M. Cobal}
\author[D]{G.-F. Dalla Betta}
\author[E]{C. Da Vi\`a}
\author[F]{G. Darbo}
\author[A]{C. Gallrapp}
\author[F]{C. Gemme}
\author[G]{F. Huegging}
\author[G]{J. Janssen}
\author[C]{A. Micelli}
\author[A]{H. Pernegger}
\author[D]{M. Povoli}
\author[G]{N. Wermes}
\author[B]{N. Zorzi}

\address[A]{CERN, Geneva 23, CH-1211, Switzerland}
\address[B]{Fondazione Bruno Kessler, FBK-CMM, Via Sommarive 18,  I-38123 Trento, Italy}
\address[C]{Universit\`a degli Studi di Udine and INFN Trieste, Gruppo Collegato di Udine, Via delle Scienze 208, I-33100 Udine, Italy}
\address[D]{DISI, Universit\`a degli Studi di Trento and INFN Padova, Gruppo Collegato d Trento, Via Sommarive 14, I-38123 Trento, Italy}
\address[E]{School of Physics and Astronomy, University of Manchester, Oxford Road, Manchester, M13 9PL, United Kingdom }
\address[F]{INFN Sezione di Genova, Via Dodecaneso 33, I-14146 Genova, Italy}
\address[G]{Physikalisches Institut, Universit$\ddot{a}$t Bonn, Nu\ss{}allee 12, D-53115 Bonn, Germany}

\cortext[cor1]{Corresponding author, now at Section de Physique (DPNC), Universit\'e\\ de Gen\`eve, 24 quai Ernest Ansermet 1211 Gen\`eve 4, Switzerland.}

%%%%%%%%%%%%%%%%%%%%%%%%%%%%%%%%%%%%%%%%%%%%%%%%%%%%%%

\begin{abstract}
In this paper we discuss results relevant to 3D Double-Side Double Type Column (3D-DDTC) pixel sensors fabricated at FBK (Trento, Italy) and oriented to the ATLAS upgrade. Some assemblies of these sensors featuring different columnar electrode configurations (2, 3, or 4 columns per pixel) and coupled to the ATLAS FEI3 read-out chip were irradiated up to large proton fluences and tested in laboratory with radioactive sources. In spite of the non optimized columnar electrode overlap, sensors exhibit reasonably good charge collection properties up to an irradiation fluence of 2\,x\,$10^{15}$\,n$_{\mathrm{eq}}$cm$^{-2}$, while requiring bias voltages in the order of 100\,V. Sensor operation is further investigated by means of TCAD simulations which can effectively explain the basic mechanisms responsible for charge loss after irradiation.  
\end{abstract}

\end{frontmatter}

\input{intro}

\input{exp}

\input{lab}

\input{tcad}
\input{conclusion}

\section{Acknowledgement}
This work has been supported in part by Provincia Autonoma di Trento within Project MEMS2, and in part by the Italian National Institute for Nuclear Physics (INFN) within Projects TREDI (CSN5) and ATLAS (CSN1). We would like to thank G. Gariano, A. Rovani and E. Ruscino (INFN - Genova), F. Rivero (Torino University) for their precious help in system assembly and measurements; R. Beccherle (INFN Genova) for designing bump bonding mask; S. Di Gioia (Selex SI) for bump bonding process. We would like also to thank: A. Dierlamm (KIT) for proton irradiation in Karlsruhe (Germany), and M. Glaser (CERN) for proton irradiation at CERN PS (CH).

\end{document}

%% file: intro.tex
\section{Introduction}
\label{sec:Intro}

The fast increase in luminosity in the modern High Energy Physics (HEP) experiments is pushing the research in the field of silicon radiation detectors to new challenging frontiers. Due to the high radiation doses foreseen for the inner tracking layers, radiation hard detectors must be designed and tested in order to provide reliable particle detection up to fluences in the order of $10^{16}$\,1-MeV equivalent neutrons per square centimeter (n$_{\mathrm{eq}}$cm$^{-2}$). 
At the same time these devices must be fast in term of charge collection time and less power consuming than the older ones. 
For these reasons several R\&D projects in the field of silicon radiation detectors have been launched in the past years, mostly focusing on the upgrades of the experiments at the Large Hadron Collider (LHC) at CERN, Geneva, Switzerland \cite{Gianotti}.\\
The main macroscopic consequences of radiation-induced defects in the detector bulk are: (i) changes in effective doping concentration, mainly with introduction of acceptor-like defects that lead to an increase of the full depletion voltage, (ii) higher leakage currents due to the creation of generation/recombination centers, and (iii) decrease of the charge collection efficiency due to carrier trapping \cite{Lindstrom}. 
The overall consequence of this damage is a strong reduction in the signal to noise ratio that can severely reduce the tracking capabilities. To counteract these effects different strategies are possible \cite{RD50}: i) material engineering, i.e., using as a substrate either non standard silicon (e.g., Magnetic Czochralski, epitaxial, etc.) or diamond, which are intrinsically more resistant to radiation damage; ii) device engineering, which consists in designing detectors with geometrical configurations that allow for lower signal degradation after irradiation. One of the most promising approaches to achieve radiation hard silicon detectors is the so-called Ò3D-architectureÓ proposed by Parker and collaborators in the mid '90s \cite{Parker}. 
In 3D detectors the electrodes have a columnar shape and are etched perpendicularly to the wafer surface, penetrating the entire sensor thickness. 
In this way the distance between electrodes is not bound to the thickness of the wafer (which is the case for standard planar silicon detectors) but can be optimized to suit performance requirements. 
Thanks to this characteristic the distance between the electrodes is decoupled from the active volumes thickness.
As a consequence, low operating voltages (less than 10\,V before irradiation, at most 200\,V after irradiation), fast response times, and strong reduction of charge trapping effects after irradiation are obtained \cite{DaVia}. 
Another important feature deriving from 3D detector technology is the active edge, which consists of a trench electrode termination allowing for a good sensitivity up to a few microns aways from the physical edge of the sensors. As a result a more efficient area coverege on wide surfaces and lower material budget are obtained \cite{Kenney02}. While active edge is an intrinsic option for 3D detectors, it can also be implemented in planar sensors, although with a major process complication  \cite{Kenney01}.
Besides all these advantages, 3D detectors have some disadvantages: in particular, the fabrication process is more complicated than a standard silicon detector process, the capacitance is higher and their response is not completely uniform because of the electrodes, that are not fully efficient, and of the presence of some low field regions. 
In order to develop 3D silicon detectors for the ATLAS upgrade the so-called ÒATLAS 3D sensor collaborationÓ \cite{PPS} was formed, involving many research centers and institutes from all over the world. 
Among the technological approaches considered for 3D fabrication, besides the original one developed at Stanford \cite{Kenney03}, there are also simplified architectural implementations. One of them, relevant to the detectors considered in this work, is the so-called 3D ÒDouble sided Double Type ColumnÓ (3D-DDTC) concept, independently proposed by FBK, Trento, Italy \cite{Zoboli} and by CNM, Barcelona, Spain \cite{Pellegrini} with the aim of reducing process complexity in view of medium volume productions. 
One of the main advantages of this approach is that it does not use a support wafer, thus avoiding the related steps of wafer bonding and final wafer removal.
Moreover, in 3D-DDTC detectors the substrate bias can be applied from the back side, making these sensors compatible with standard planar sensors and easing the detector assembly within a tracking system.
Columns are etched from both wafer sides (n$^{+}$ from the top, p$^{+}$ from the bottom) and do not pass through the entire wafer thickness, so they only partially overlap. 
From TCAD simulations \cite{Zoboli}, it was predicted that the performance of 3D-DDTC detectors is comparable to that of standard 3D detectors if the column overlap is a significant fraction of the wafer thickness, whereas it can be degraded if column thicknesses are not optimized, which is the case for the first prototypes fabricated at FBK and considered in this paper. 
Therefore, the radiation hardness should be carefully studied in order to obtain useful information for the design and technology optimization. The FBK devices were previously tested both in laboratory \cite{DallaBetta} and in beam tests at CERN in pre-irradiation conditions \cite{Grenier}, obtaining very good results. In order to study their radiation hardness, different irradiation campaigns were conducted, and irradiated detectors were measured again in a test beam at CERN \cite{Micelli} and in laboratory. \\
In this paper we report on selected results from functional characterization with radioactive sources conducted in laboratory on these 3D-DDTC detectors. Numerical simulations are also used to gain better insight into experimental results. The paper is organized as follows: Section 2 gives a brief description of devices under test and summarizes the two proton irradiation campaigns; Section 3 describes and discusses post-irradiation measurement results also comparing them with pre-irradiation results; Section 4 reports numerical simulation results and compares them to the measurements. Conclusions follow.

%% file: exp.tex
\section{Experimental}
\label{sec:exp}

\subsection{Sensor description}
\begin{figure}[htb]
\centering
\includegraphics[scale=0.75]{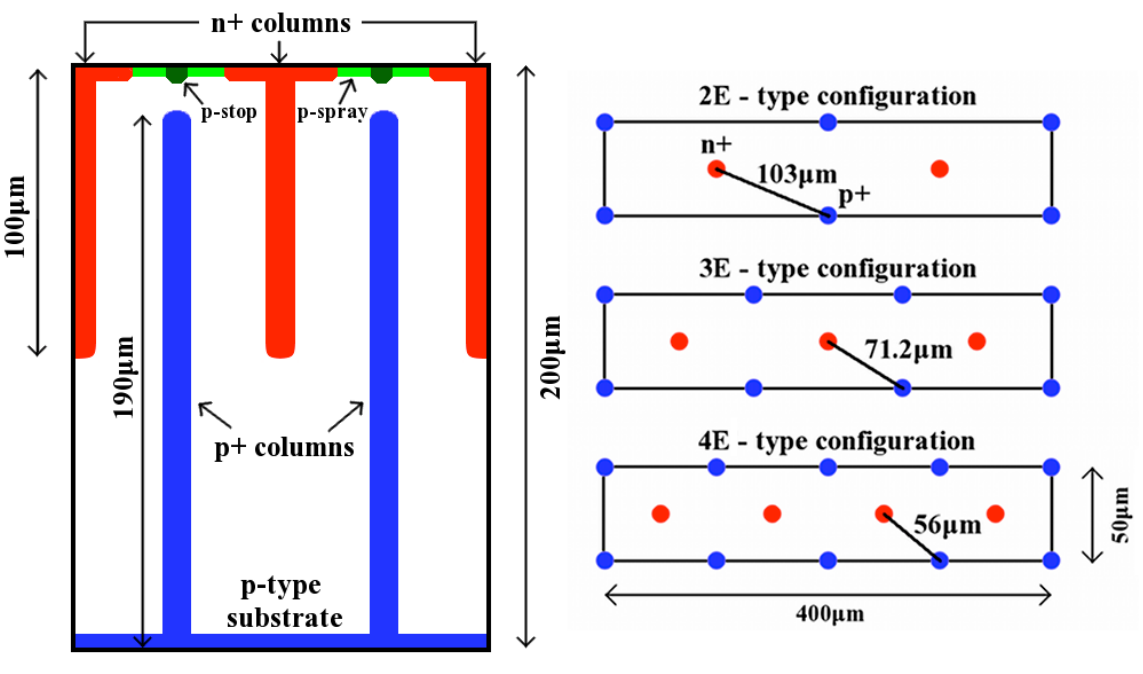}
\caption{(left) Schematic cross-section of the sensors and (right) different pixel configurations.}
\label{fig:F1}
\end{figure}

The sensors under test are 3D-DDTC detectors fabricated at FBK on 4", 200\,$\mu$m thick, p-type, FZ silicon wafers. As previously stated, in these devices columns are etched from opposite sides of the wafer and are not completely passing through the silicon bulk \cite{Kenney03}. At the time of fabrication of these sensors (2008) the Deep Reactive Ion Etching (DRIE) equipment was not yet available at FBK so the etching of the holes was commissioned to an external company (IBS, France) and problems related to the calibration of this step led to an un-even column depth  (see Fig.\,\ref{fig:F1} (left)). This problem translated into a relatively small column overlap, in the order of 90\,$\mu$m, which of course is not ideal and could affect the device performance, especially after irradiation. 
The nominal column diameter is 10\,$\mu$m. The surface insulation between n$^{+}$ electrodes on the front side is achieved by combined p-spray and p-stop implants \cite{Piemonte}, whereas all ohmic columns are shorted on the back side by uniform p$^{+}$ diffusion and metal.\\
Devices under test are pixel detectors compatible with the FEI3 ATLAS read-out chip \cite{Peric}, and feature various layout options differing in the number of columns per pixel: 2E-type (two n$^{+}$ columns per pixel), 3E-type (three n$^{+}$  columns per pixel) and 4E-type (four n$^{+}$  columns per pixel). The number of ohmic columns per pixel also changes accordingly, and the inter-electrode distances are 103\,$\mu$m, 71.2\,$\mu$m, and 56\,$\mu$m, respectively (see Fig.\,\ref{fig:F1} (right)). 
Detectors were bump bonded to FEI3 ATLAS read-out chips at SELEX \cite{SELEX}, Rome, Italy. The FEI3 chip was designed with radiation tolerant layout rules in a 0.25\,$\mu$m CMOS technology. The maximum radiation dose that the chip can withstand is in the order of 50\,Mrad. 
Detectors were designed to exactly match the geometry of the readout channels of the chip. The front-end chip presents 2880 channels which are arranged in a matrix of 160 rows per 18 columns, with a pixel size of  400\,$\mu$m\,x\,50\,$\mu$m. Each channel is composed by an analog and a digital part. The analog part integrates the sensor output current by means of a charge sensitive preamplifier with constant current discharge, thus yielding a triangular pulse shape, which is fed to a discriminator along with a pre-set threshold. As a result, the width of the discriminator output signal, i.e., the Time Over Threshold (TOT) expressed in units of 40\,MHz clock, is in first approximation proportional to the collected charge. A detailed explanation of the FEI3 operation can be found in \cite{Peric}.

\subsection{Proton irradiation campaigns}
In order to study the radiation tolerance of these sensors, five devices were irradiated to two different proton fluences at two facilities: three samples at the Karlsruhe Institute of Technology (KIT) with a 25\,MeV proton beam up to 1\,x\,10$^{15}$n$_{\mathrm{eq}}$cm$^{-2}$ and two samples at CERN PS with a 24\,GeV/c proton beam up to 2\,x\,10$^{15}$n$_{\mathrm{eq}}$cm$^{-2}$ (see Table\,\ref{tab:T1}).  The proton fluences were scaled to 1\,MeV equivalent neutrons per square centimeter (n$_{\mathrm{eq}}$cm$^{-2}$) using the NIEL hypothesis with hardness factors of 1.85 and 0.62 for 25\,MeV and 24\,GeV protons, respectively. The uncertainty in the irradiation fluences is lower than 10\,\%. After the irradiation, the detectors were cooled to prevent annealing. However, the detectors had to be kept at room temperature for a short time during handling and measurement setup, so that they experienced some annealing (which was of fractions of an hour).
\begin{table*}[htb]
\begin{center}
\begin{tabular}{|p{2cm}|*{5}{c|}}
\hline
Module ID & Sensor Type& Fluence [n$_{\mathrm{eq}}$cm$^{-2}$] & Particle & Facility \\
\hline
A	&	2E	&	1x$10^{15}$	&	25\,MeV proton		&	KIT	\\
B	&	3E	&	1x$10^{15}$	&	25\,MeV proton		&	KIT	\\
C	&	4E	&	1x$10^{15}$	&	25\,MeV proton		&	KIT	\\
D	&	2E	&	2x$10^{15}$	&	24\,GeV proton		&	CERN PS	\\
E	&	4E	&	2x$10^{15}$	&	24\,GeV proton		&	CERN PS	\\
\hline
\end{tabular}
\end{center}
\caption{Overview of irradiated assemblies. }
\label{tab:T1}
\end{table*}

%% file: lab.tex
\section{Functional lab-test measurements}
\label{sec:lab}

\begin{figure}[htb]
\centering
\includegraphics[scale=0.45]{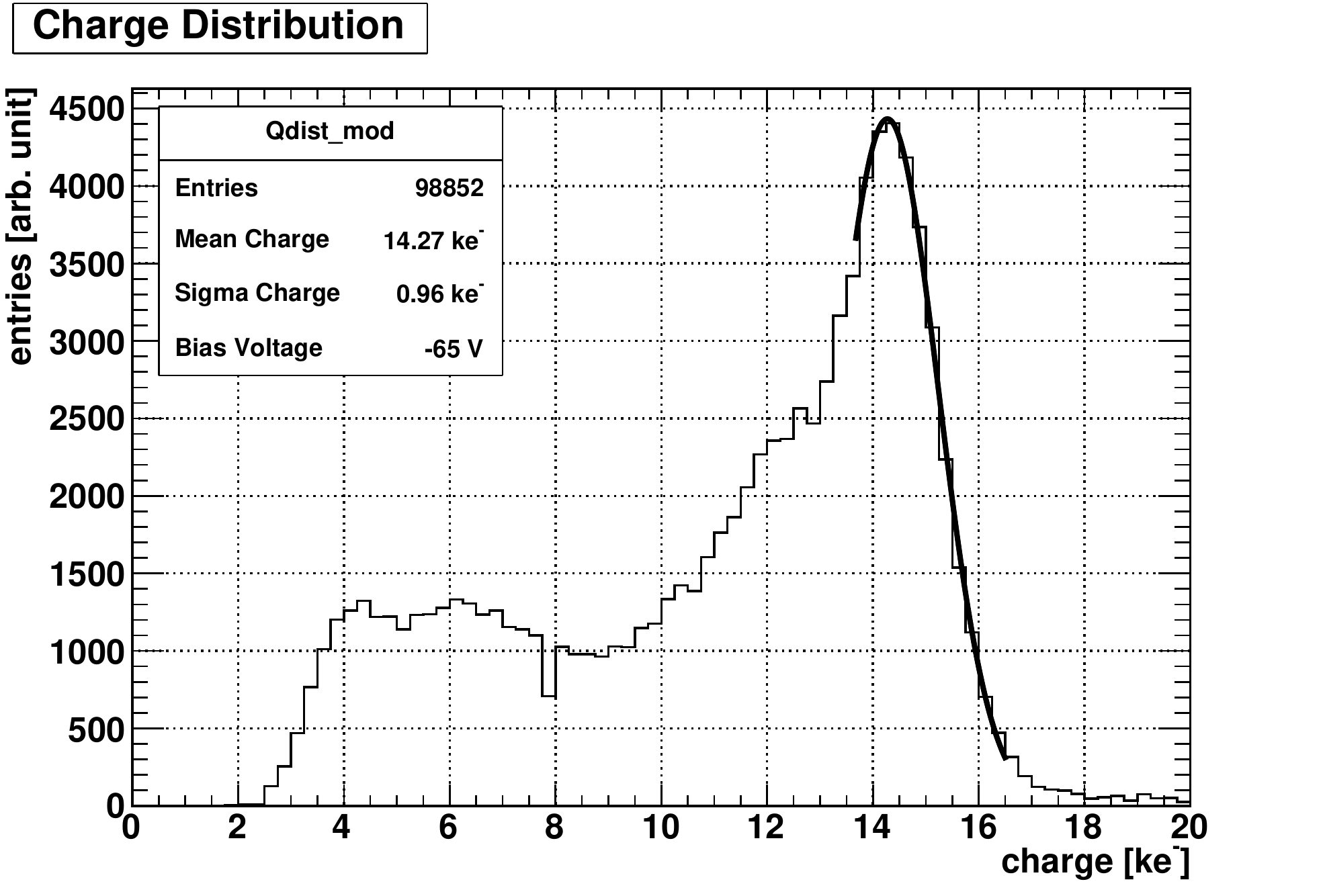}
\caption{$^{241}$Am spectrum measured with a 4E-type detector (sample C) irradiated with protons at 1\,x\,10$^{15}$\,n$_{\mathrm{eq}}$cm$^{-2}$ and reverse biased at 65\,V.}
\label{fig:F5}
\end{figure}

\begin{figure}[h!]
\centering
\subfigure[]{
\includegraphics[scale=0.65]{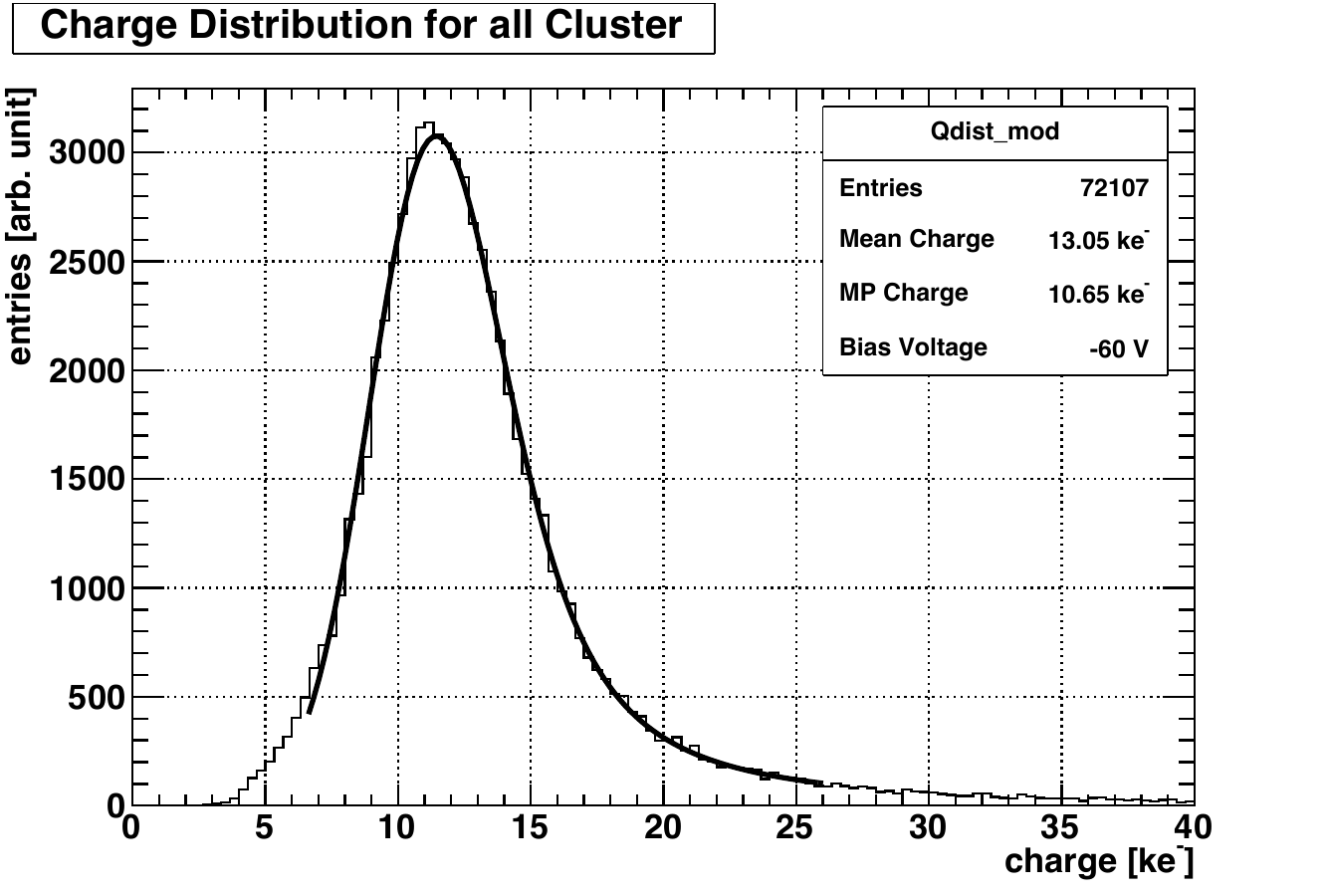}
\label{fig:F6a}
}
\subfigure[]{
\includegraphics[scale=0.44]{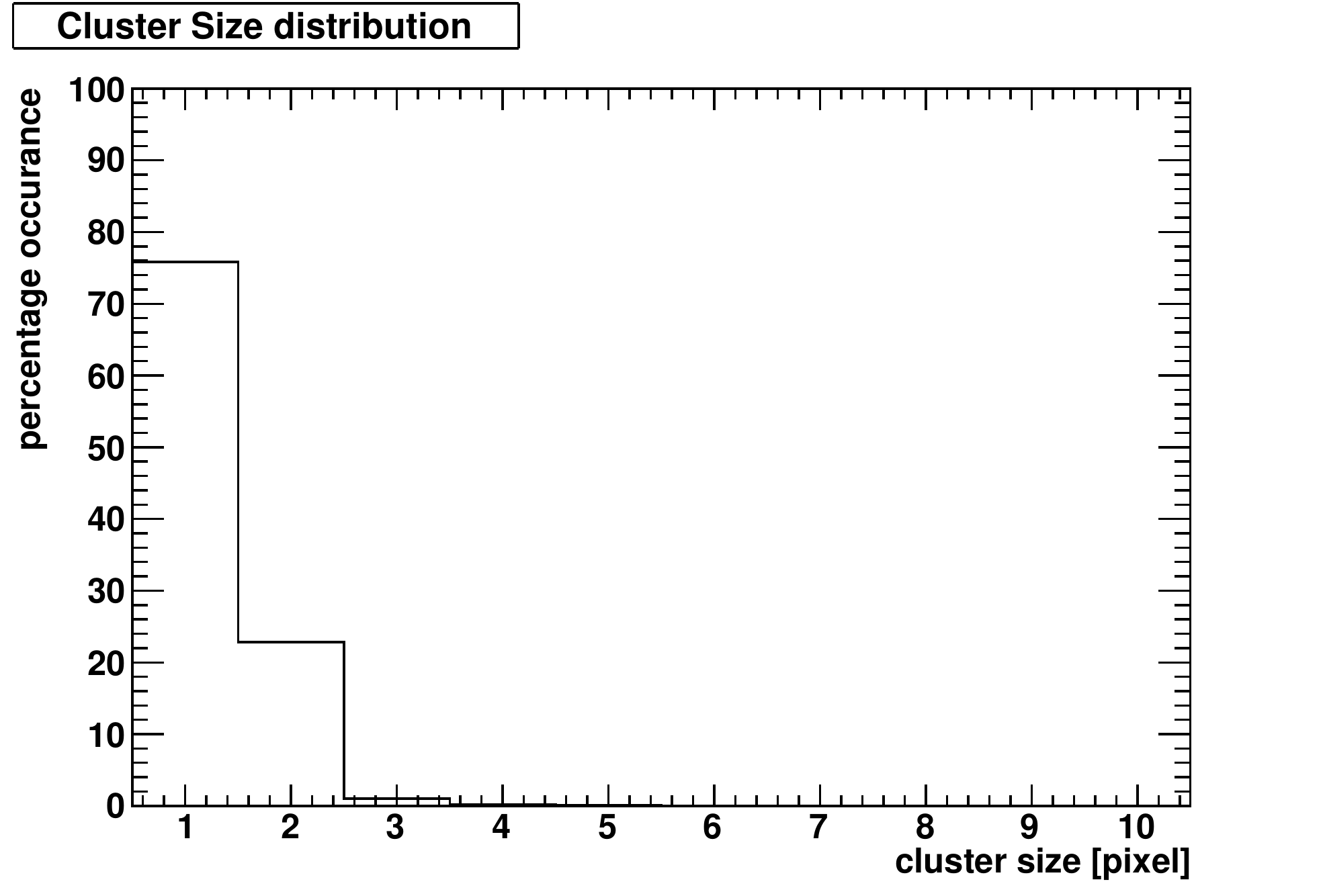}
\label{fig:F6b}
}
\caption{Results for tests with  $^{90}$Sr source on a 4E-type detector (sample C) irradiated with protons at 1\,x\,10$^{15}$\,n$_{\mathrm{eq}}$cm$^{-2}$ and biased at 60\,V: \subref{fig:F6a} pulse height distribution, and \subref{fig:F6b} cluster size distribution.}
\end{figure}

Results from the electrical and functional characterization before irradiation of pixel detectors here considered are reported in \cite{DallaBetta}, \cite{LaRosa}. 
In this paper we will focus on measurements performed on irradiated devices and results of non-irradiated detectors will be recalled for a direct comparison (see a summary in Table \ref{tab:T2}). 
The sample characterization has been carried out by measuring: leakage current versus bias voltages, threshold, noise and response to radioactive $\gamma$- and $\beta$-sources. 
All measurements on irradiated devices were performed at $-20\,^{\circ}\mathrm{C}$  in order to reduce leakage current and avoid reverse annealing effects.
\begin{figure}[h!]
\centering
\includegraphics[scale=0.45]{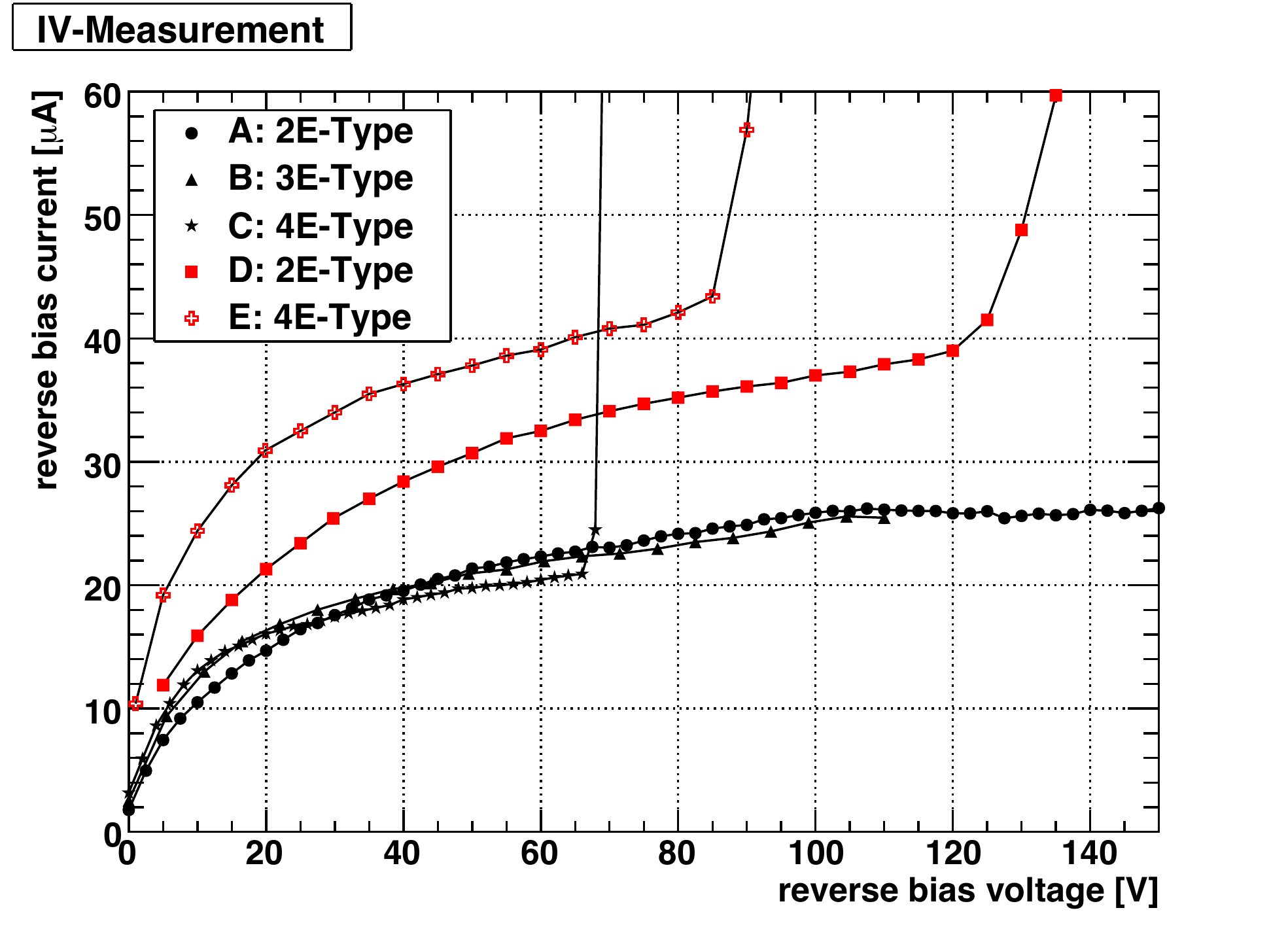}
\caption{Overview of I-V curves from irradiated assemblies.}
\label{fig:F2}
\end{figure}
\newline Before irradiation detectors had leakage currents of a few hundred nA at room temperature and breakdown voltages between 60\,V and 70\,V. 
After irradiation, the leakage currents and the breakdown voltages shift to higher values due to the displacement damage of the silicon and to the increase of the oxide charge concentration caused by ionizing radiation. It was measured that the leakage current increased by several orders of magnitude, whereas, the breakdown voltages were generally increased, sometimes well above 100\,V, although results are not completely uniform. 
Breakdown voltage improvement after irradiation is typical of p-spray isolated planar structures \cite{Piemonte}. 
These 3D-DDTC sensors confirm the expectations. Fig.\,\ref{fig:F2} shows an overview of leakage current versus bias voltage: the current values and trends are in agreement with expectations, with different current levels corresponding to different irradiation fluences. 
Assemblies A, B and C were irradiated at 1\,x\,10$^{15}$\,n$_{\mathrm{eq}}$cm$^{-2}$ and show similar current values after irradiation (reaching about 25\,$\mu$A). A clear saturation at about 100\,V can be observed for samples A and B, whereas sample C suffers from an earlier breakdown at about 65\,V. 
Assemblies D and E were irradiated at 2\,x\,10$^{15}$\,n$_{\mathrm{eq}}$cm$^{-2}$ and reach currents of 35 to 40\,$\mu$A (not yet fully saturated) before breakdown. The agreement between the leakage current damage constant ($\alpha$ $\approx$ 5\,x\,10$^{-17}$ A/cm) extracted from these measurements and the generally accepted value of 4\,x\,10$^{-17}$ A/cm \cite{Lindstrom} is good enough considering the uncertainties in the irradiation fluence, annealing conditions and temperature.\\
In order to evaluate the performance of the assemblies under test (sensor and front-end electronics), the system was calibrated aiming at a ToT of 60 units for a charge of 20\,ke and a threshold of 3.2\,ke. 
Threshold and noise measurements have been performed on each pixel based on S-curve fit function \cite{ATLAS}. 
Table \ref{tab:T3} summarizes the values of threshold and noise for the irradiated assemblies. Compared to values measured before irradiation (cf. Table \ref{tab:T2}), both threshold average and dispersion are very similar. Also noise values, expressed as equivalent noise charge (ENC), are only slightly different than before irradiation. 
\begin{figure}[htb]
\centering
\includegraphics[scale=0.34]{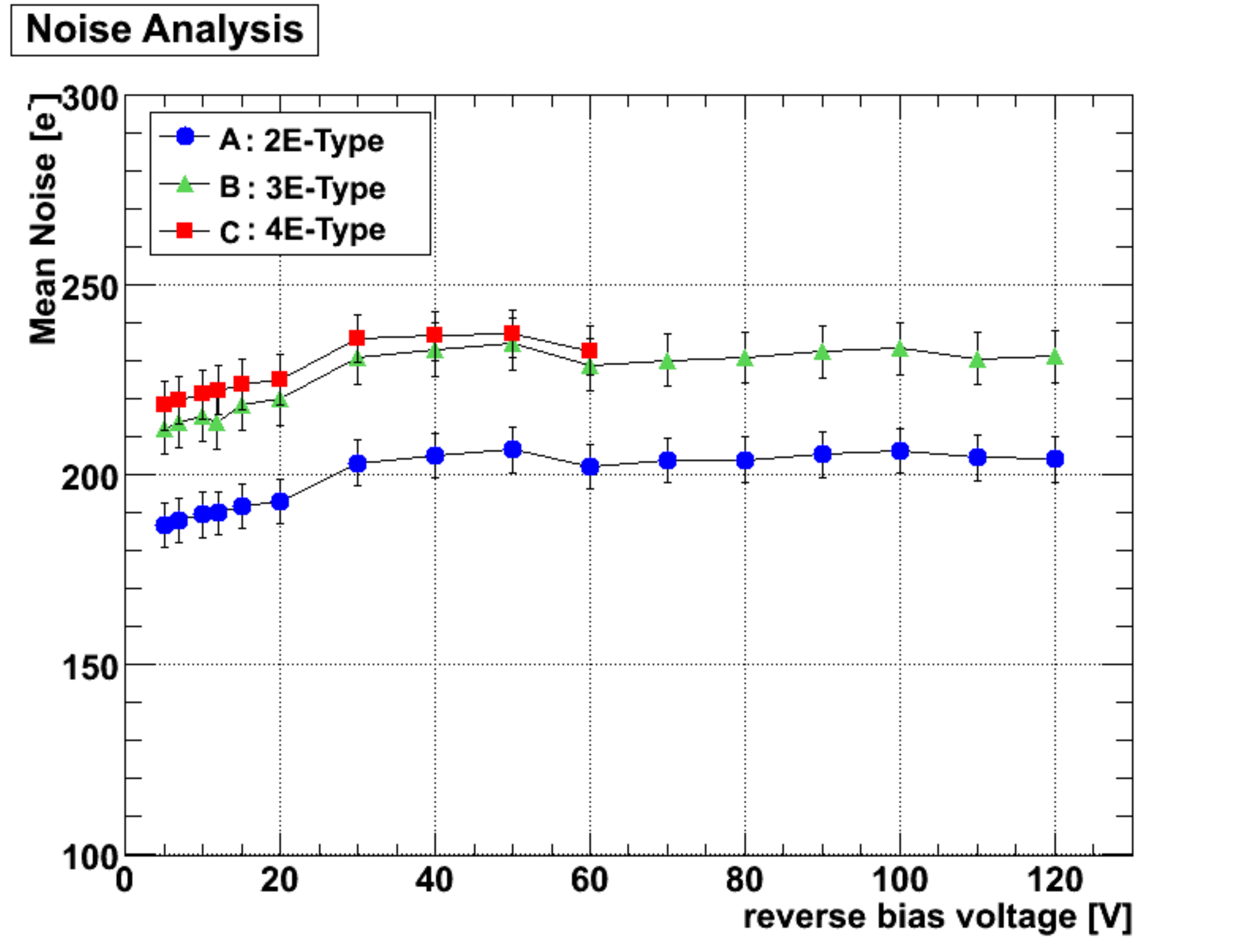}
\caption{Equivalent noise charge as a function of bias voltages for irradiated assemblies A, B, and C.}
\label{fig:F3}
\end{figure}
As an example, Fig.\,\ref{fig:F3} shows the ENC as a function of the voltage for sensors A, B, and C. The ENC curves are almost flat because, after irradiation to such a large fluence, the capacitance, which is the main factor for noise, is almost constant with bias due to the very high resistivity of the substrate \cite{Bates}. Different noise levels are observed, in agreement with the pre-irradiation case, because of the different capacitances characterizing the different column configurations (cf. Table \ref{tab:T2}). 
A direct comparison in the noise values before and after irradiation is difficult because of the different temperatures ($+20\,^{\circ}\mathrm{C}$ vs $-20\,^{\circ}\mathrm{C}$): on one hand, the increase of leakage current in irradiated sensors would justify an increase in the noise in the order of tens of electrons rms \cite{Blanquart}, on the other hand this increase might well be compensated by the much lower temperature used for measurements after irradiation. Limiting the comparison to the irradiated samples, it might be puzzling that those irradiated at 2\,x\,10$^{15}$\,n$_{\mathrm{eq}}$cm$^{-2}$ (samples D and E) show lower noise than those irradiated at 1\,x\,10$^{15}$\,n$_{\mathrm{eq}}$cm$^{-2}$ (samples A, B, and C), but it should be noted that this difference might be due to the radiation damage to the electronics rather than to the sensors: in fact, for the 24-GeV proton irradiation a Total Ionizing Dose (TID) of about 94\,Mrad has been estimated, whereas the estimated TID is 144\,Mrad for 25-MeV proton irradiation. 
Considering that the letter TID is about 35\% higher than the former and about three times the one for which the FEI3 chip has been designed (50\,Mrad), it could be responsible for a degradation in the chip noise performance.\\
Charge collection mechanisms were studied by means of an $^{241}$Am $\gamma$-source and a $^{90}$Sr $\beta$-source, comparing the results with those obtained before irradiation (cf. Table \ref{tab:T2}). For $^{241}$Am measurements, the self-triggering capabilities of the system were exploited, whereas for $^{90}$Sr measurements the trigger was taken from a scintillator placed behind the devices under test.\\
Table\,\ref{tab:T4} summarizes the results of $^{241}$Am $\gamma$-source measurements for the three sensors irradiated at 1\,x\,10$^{15}$\,n$_{\mathrm{eq}}$cm$^{-2}$ (data are not available for samples D and E). 
\begin{figure}[htb]
\centering
\includegraphics[scale=0.46]{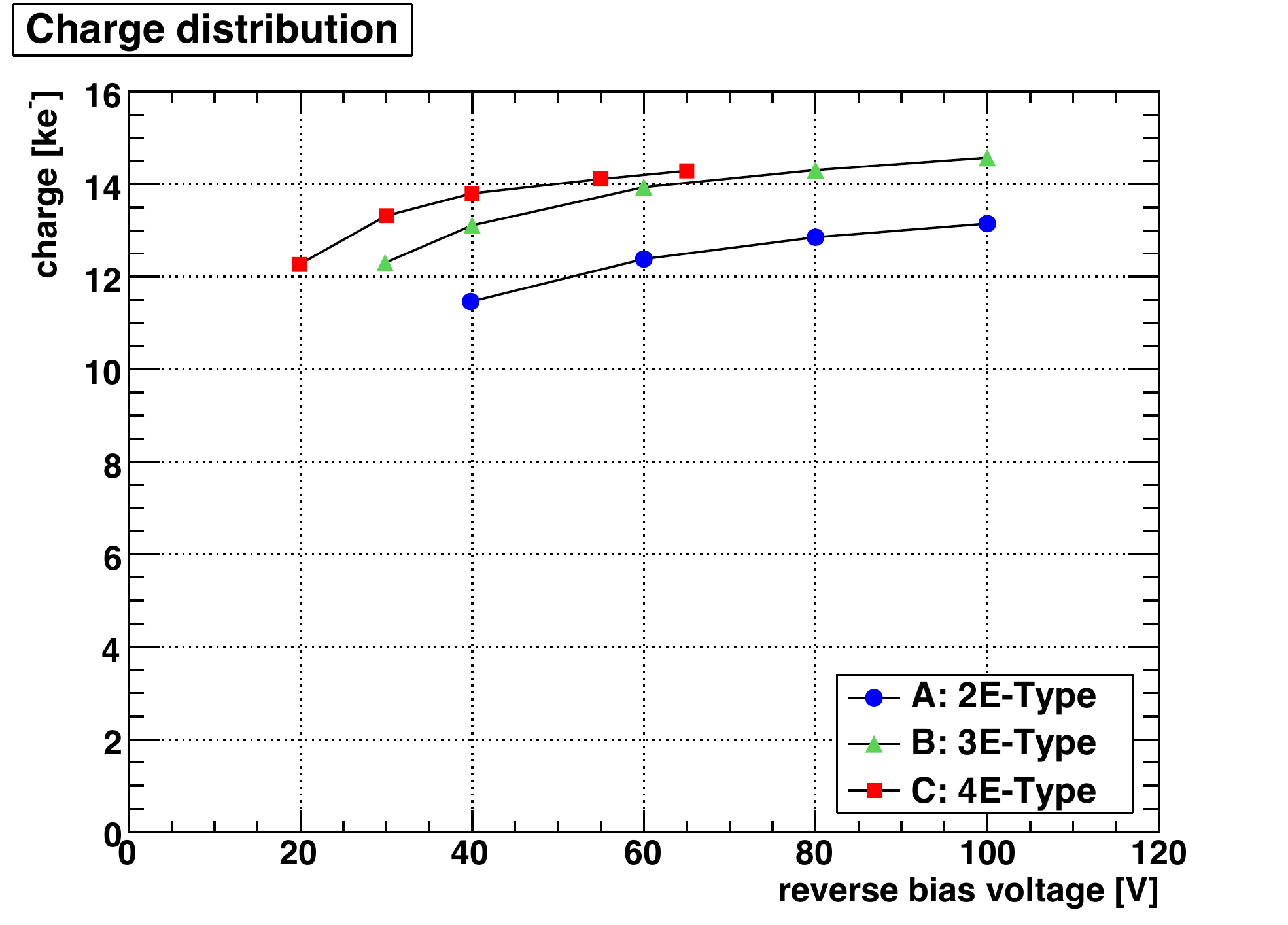}
\caption{Mean peak (60-keV) of the collected charge for $^{241}$Am source measurements as a function of reverse bias voltage in all irradiated samples. }
\label{fig:F4}
\end{figure}
Before irradiation (cf. Table \ref{tab:T2}) the mean of the fitted Gaussian of about 14.5\,ke without clustering had been obtained with an applied reverse bias of 35\,V. This charge corresponds to the energy peak at 60\,keV, in agreement with theoretical expectations within the uncertainty due to the calibration process, which was estimated to be in the order of 10-15\,\%. 
This uncertainty  is due to the Time over Threshold (ToT) to charge calibration process which contains several components and it estimated to be about mentioned percentage.  In a first approximation, the 10-15\,\%  of uncertainty is related to the uncertainty of the capacitance used to tune the first stage of the FE in the calibration process.  Detailed explanation can be found in \cite{Peric}.\\
Looking at Table\,\ref{tab:T4}, it is possible to notice that after irradiation, samples B and C yield a value of collected charge comparable to the value before irradiation, whereas sample A (2E-type) shows lower collected charge. This can be ascribed to the longer distance between the n$^{+}$ and p$^{+}$ electrodes in sample A, which results in higher trapping probability. The importance of the column geometry is confirmed by results in Fig.\,\ref{fig:F4}, which shows the mean charge values as a function of the bias voltage for the three samples. Sensor C (4E-type) reaches roughly the same maximum charge as sensor B (3E-type), but at a lower voltage, owing to a lower distance between electrodes. Even higher voltage is necessary to have a significant charge collection in sensor A, which also tends to saturate at a lower charge value. An example of $^{241}$Am spectrum measured with sample C is shown in Fig.\,\ref{fig:F5}. 
The main peak corresponding to the 60-keV photons is distorted with a tail towards lower charge values, that can be attributed to the absorption of photons in regions with lower electric field, for which the charge collection is not fully efficient.\\
Table \ref{tab:T5} summarizes $\beta$-source measurements for all irradiated detectors biased at the optimal voltage. The devices have been measured at different bias voltages, always stopping at a voltage lower than the breakdown voltage.
Fig.\,\ref{fig:F6a} shows the pulse height spectrum in response to a $^{90}$Sr $\beta$-source in the 4E sensor irradiated with protons at 1\,x\,10$^{15}$\,n$_{\mathrm{eq}}$cm$^{-2}$ (sample C). The distribution is related to all clusters, and charge values in excess of 7\,ke have been fitted with a Landau function, which is also shown in the figure. The most probable value (MPV) of the collected charge is 10.65\,ke, whereas it was about 15\,ke before irradiation (cf. Table\,\ref{tab:T2}). Thus, the degradation in charge collection is more pronounced than that observed for the $^{241}$Am $\gamma$-source measurements. This is generally the case for all the irradiated assemblies, as can be seen from the values in Table \ref{tab:T5}. The reason for this behavior is related to the different ways charge is generated inside the sensors: while every $\gamma$-ray in silicon can be absorbed at different depths and the released charge is localized in a region of a few cubic micrometers, $\beta$ particles traverse the whole bulk thickness releasing charge along the entire track. This means that, if the device is not fully depleted, part of the charge released by a $\beta$ particle will be generated in a non depleted region and therefore almost completely lost (due to recombination). On the contrary, for all the $\gamma$-rays absorbed in the region where columns overlap the generated charge will be collected also after irradiation, except for the case where the inter-electrode distance is too large, because of trapping. Assemblies irradiated at 1\,x\,10$^{15}$\,n$_{\mathrm{eq}}$cm$^{-2}$ perform reasonably well, considering that the applied voltage is high enough to achieve lateral depletion between the columns.
However, it should be stressed that a non negligible fraction of the active volume at the bottom is not depleted, thus causing charge loss (this effect will be better explained in the following with the aid of TCAD simulations). Samples irradiated at 2\,x\,10$^{15}$\,n$_{\mathrm{eq}}$cm$^{-2}$ show further degradation of the collected charge with respect to those irradiated at 1\,x\,10$^{15}$\,n$_{\mathrm{eq}}$cm$^{-2}$, due to stronger trapping effects. Nevertheless, the collected charge is still reasonably good because of the higher bias voltage that could be applied to these sensors before they reached breakdown, which allowed lateral depletion between columns to be achieved both for the 2E and the 4E configurations. Again, the 4E sample collects more charge than the 2E one at a lower voltage owing to the shorter distance between the electrodes. 
The cluster size distribution for sample C is also shown in Fig.\,\ref{fig:F6b}. Most events ($\sim$80\,\%) are cluster size 1, the remaining part ($\sim$20\,\%) are cluster size 2. This distribution can be explained taking into account that $\beta$ particles reaching the sensor are not all perpendicular to the surface due to the geometry of the collimator. The angle distribution was found to span from $90\,^{\circ}$ to $84.29\,^{\circ}$. Although in 3D sensors the electric field distribution provides a sort of self-shielding effect in each pixel, in case particle hits are not perpendicular to the surface the probability of charge sharing between two adjacent pixels becomes non negligible.\\
The performances of some of the irradiated devices considered in this paper were also tested with a 120 GeV/c $\pi^{+}$ beam at CERN SPS in June 2010. The aim of this test was to study tracking efficiency, charge sharing and cluster size. A detailed description of the obtained results can be found in \cite{Micelli}. It should be noticed that both the values of the collected charge and the cluster size distributions are in very good agreement with those measured in laboratory.\\
The MPV of the collected charge reported in Table \ref{tab:T5} are relevant to the highest possible bias voltage. The variation of the MPV with the bias voltage is also of interest to better understand the behavior of these sensors. 
\begin{figure}[htb]
\centering
\includegraphics[scale=0.45]{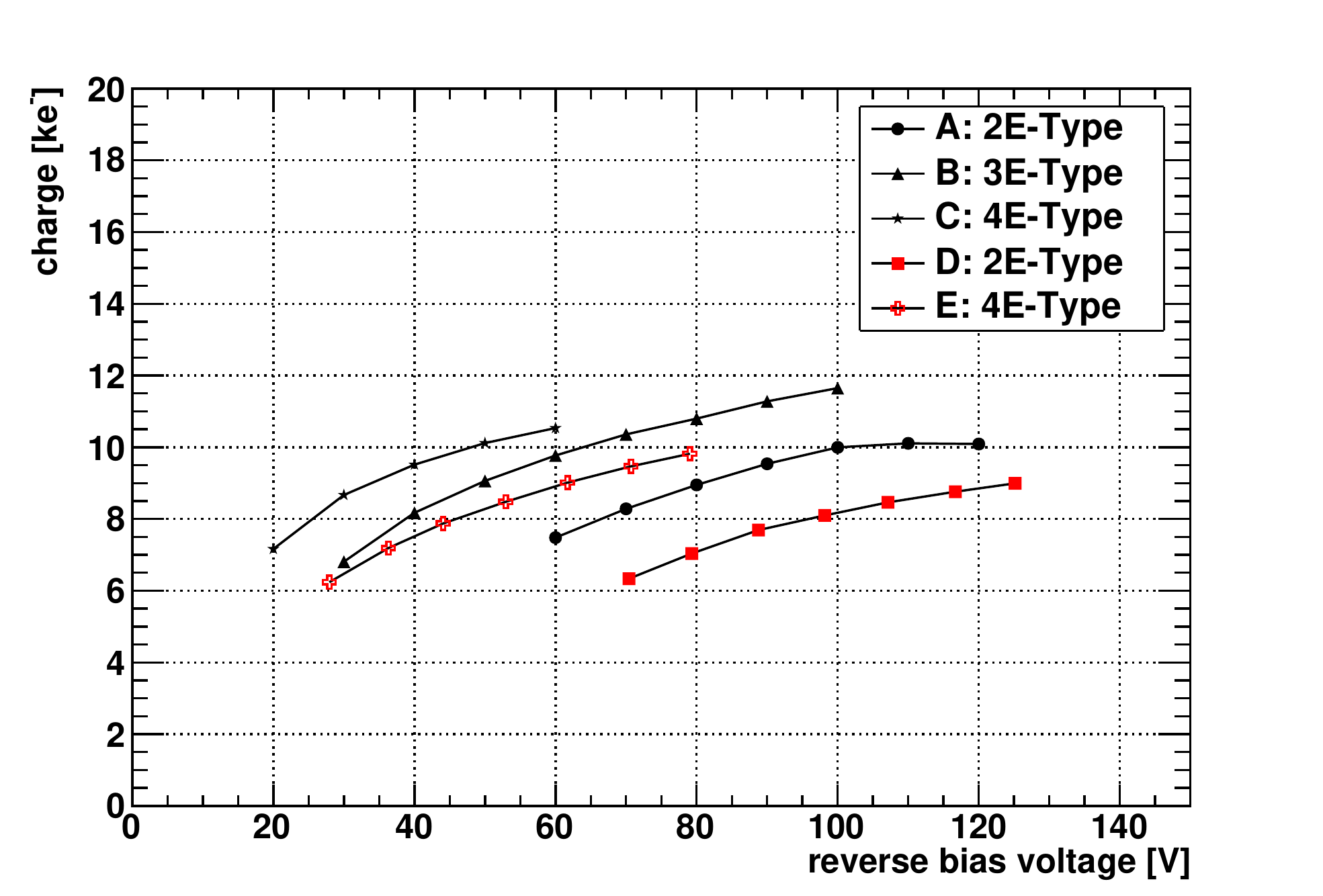}
\caption{Most Probable Value (MPV) of the collected charge for $^{90}$Sr source tests as a function of reverse bias voltage in all irradiated samples.}
\label{fig:F7}
\end{figure}
Fig.\,\ref{fig:F7} summarizes data for all the irradiated assemblies. As expected, the collected charge increases with the applied voltage in agreement with the larger depleted volume within the sensors. It should also mentioned that a loss of efficiency associated to the increasing number of electrodes was measured in the CERN SPS beam-tests as is reported in \cite{Micelli}.
Apart from sensor A, for which trapping effects are more severe as also observed with $\gamma$-rays, charge values are not yet completely saturated in the considered voltage range. 
As expected, at the same bias voltage the collected charge increases with the number of electrodes. The same value of collected charge is observed at a lower voltage in 4E samples with respect to 3E and 2E ones. Samples irradiated at 2\,x\,10$^{15}$\,n$_{\mathrm{eq}}$cm$^{-2}$ of course need a higher bias voltage to reach a sufficient charge collection efficiency, and maximum values are slightly lower due to trapping.
%
%
%-------- TABLES
%
%
\begin{table*}[htb]
\begin{center}
\begin{tabular}{|p{2cm}|*{7}{c|}}
\hline
Sensor type & I$_{\mathrm{leakage}}$ [nA] & V$_{\mathrm{breakdown}}$ [V] & Threshold [e]	& ENC [e]	& $^{241}$Am 60-keV peak [ke]	& $^{90}$Sr MPV  peak [ke] \\
\hline
2E	&	250\,-\,300	&	60\,-\,70	&	3240\,$\pm$\,54	&	197\,$\pm$\,9	&	14.71	&	15.35	\\
3E	&	250\,-\,300	&	60\,-\,70	&	3274\,$\pm$\,51	&	206\,$\pm$\,8	&	14.50	&	15.25	\\
4E	&	250\,-\,300	&	60\,-\,70	&	3297\,$\pm$\,56	&	227\,$\pm$\,8	&	14.37	&	15.25	\\
\hline
\end{tabular}
\end{center}
\caption{Summary of typical results obtained from non-irradiated sensors. Except for V$_{\mathrm{breakdown}}$, the parameters are obtained at a bias voltage of 35\,V.}
\label{tab:T2}
\end{table*}
\begin{table*}[htb]
\begin{center}
\begin{tabular}{|p{2cm}|*{5}{c|}}
\hline
Module ID 	&	Sensor type	 &	Threshold [e]	&	ENC [e]	&	V$_{\mathrm{test}}$ [V]\\
\hline
A	&	2E	&	3261\,$\pm$\,73	&	204.1\,$\pm$\,11.6	&	120 \\		
B	&	3E	&	3158\,$\pm$\,140	&	233.2\,$\pm$\,13.4	&	100 \\		
C	&	4E	&	3267\,$\pm$\,78	&	232.4\,$\pm$\,12.7	&	60 \\	
D	&	2E	&	2950\,$\pm$\,206	&	119.4\,$\pm$\,31.5	&	125 \\	
E	&	4E	&	3307\,$\pm$\,115	&	189.6\,$\pm$\,19.0	&	80 \\		
\hline
\end{tabular}
\end{center}
\caption{Threshold and noise values measured from irradiated assemblies.}
\label{tab:T3}
\end{table*}
\begin{table*}[htb]
\begin{center}
\begin{tabular}{|p{2cm}|*{4}{c|}}
\hline
Module ID 	&	Sensor type	 &	$^{241}$Am 60-keV peak [ke]	&	V$_{\mathrm{test}}$ [V]\\
\hline
A	&	2E	&	12.8	&	80 \\		
B	&	3E	&	14.5	&	100 \\		
C	&	4E	&	14.3	&	65 \\			
\hline
\end{tabular}
\end{center}
\caption{Collected charge from the assemblies irradiated at 1\,x\,10$^{15}$\,n$_{\mathrm{eq}}$cm$^{-2}$ exposed to $^{241}$Am source.}
\label{tab:T4}
\end{table*}
\begin{table*}[htb]
\begin{center}
\begin{tabular}{|p{2cm}|*{4}{c|}}
\hline
Module ID 	&	Sensor type	 &	$^{90}$Sr MPV peak [ke]	&	V$_{\mathrm{test}}$ [V]\\
\hline
A	&	2E	&	10.2	&	120 \\		
B	&	3E	&	11.9	&	100 \\		
C	&	4E	&	10.9	&	60 \\	
D	&	4E	&	9.0	&	125 \\		
E	&	4E	&	9.8	&	80 \\				
\hline
\end{tabular}
\end{center}
\caption{Collected charge from all irradiated assemblies exposed to $^{90}$Sr source.}
\label{tab:T5}
\end{table*}

%% file: tcad.tex
\section{TCAD simulations}
\label{sec:tcad}

% 1\,x\,10$^{15}$\,n$_{\mathrm{eq}}$cm$^{-2}$

% $^{90}$Sr $\beta$-source

% $^{241}$Am $\gamma$-source

% $+20\,^{\circ}\mathrm{C}$ 

\begin{figure}[t]
\centering
\includegraphics[scale=0.72]{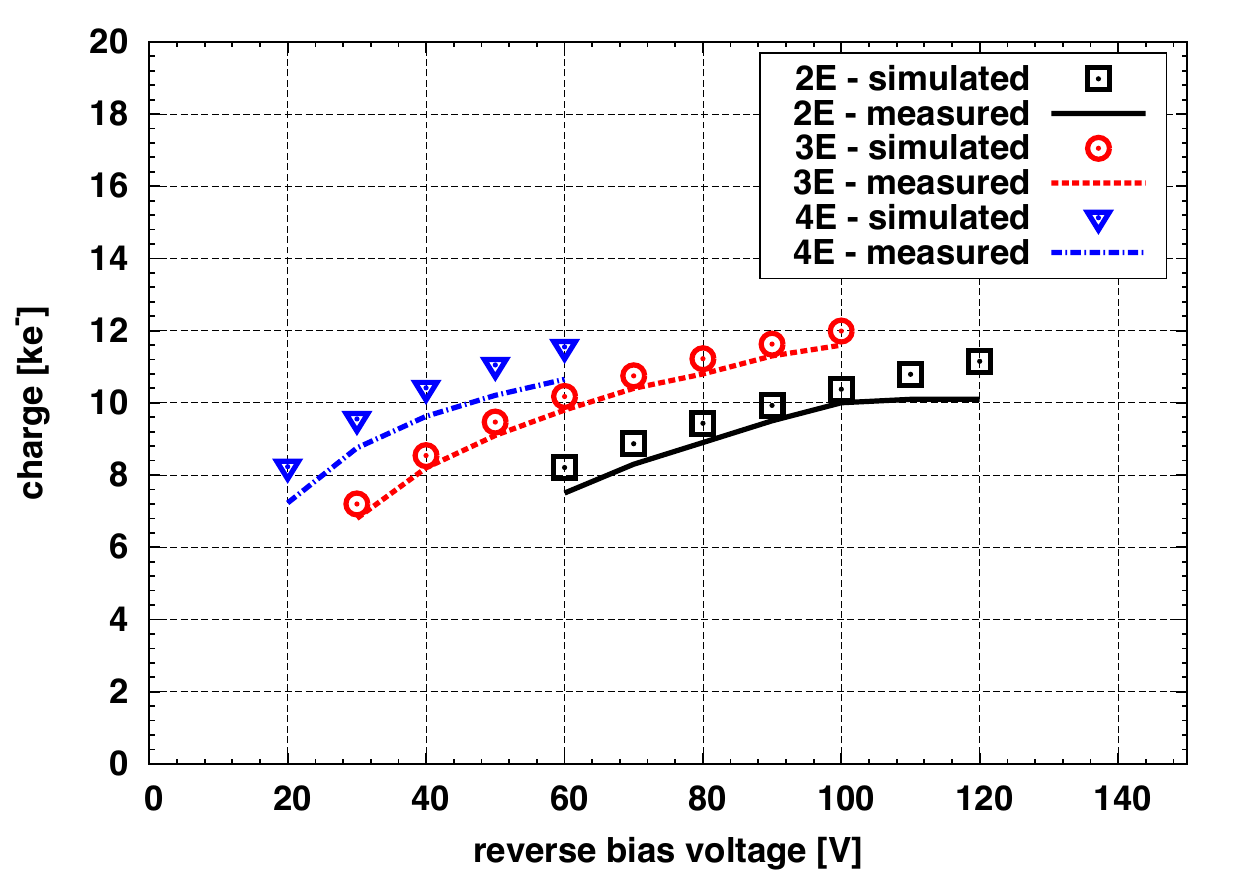}
\caption{Comparison between measured and simulated collected charge as a function of reverse bias voltage for sensors at 1\,x\,10$^{15}$\,n$_{\mathrm{eq}}$cm$^{-2}$.}
\label{fig:F9}
\end{figure}

\begin{figure*}[h!]
\centering
\includegraphics[scale=0.3]{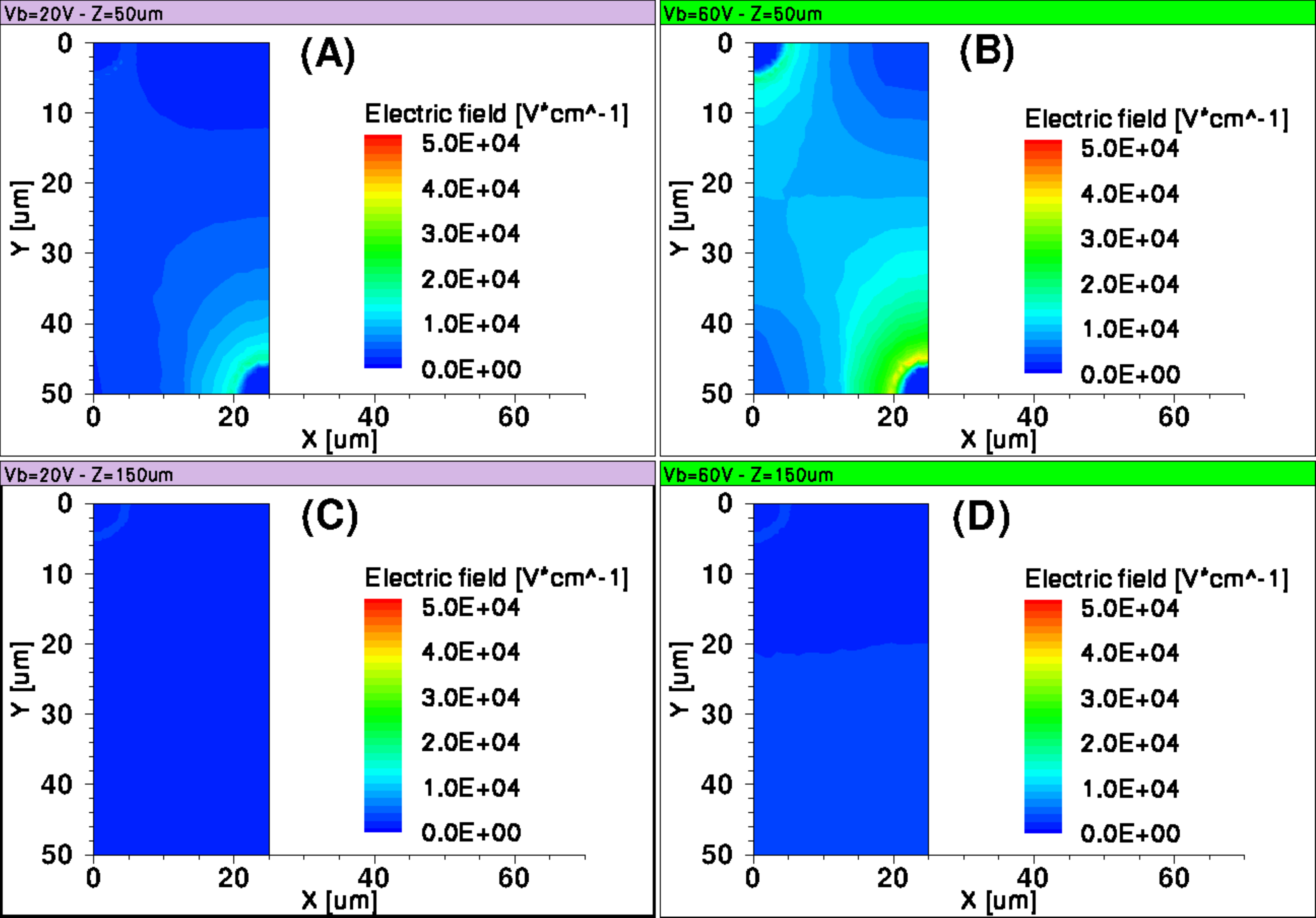}
\caption{2d cuts in the 4E sensor showing the electric field at two different depths (Z) and two bias voltages (V$_{\mathrm{b}}$): (A) Z\,=\,50\,$\mu$m, V$_{\mathrm{b}}$\,=\,20\,V;  (B) Z\,=\,50\,$\mu$m, V$_{\mathrm{b}}$\,=\,60\,V;  (C) Z\,=\,150\,$\mu$m, V$_{\mathrm{b}}$\,=\,20\,V; (D) Z\,=\,150\,$\mu$m, V$_{\mathrm{b}}$\,=\,60\,V.}
\label{fig:F10}
\end{figure*}

In order to better understand the experimental results, numerical simulations were performed with TCAD tools from Synopsys \cite{Synopsys}. Simulations were focused on devices measured with $^{90}$Sr $\beta$-source and irradiated at at 1\,x\,10$^{15}$\,n$_{\mathrm{eq}}$cm$^{-2}$, in order to limit the computational load and reduce the simulation time.\\
Simulations were performed at different bias voltages. Moreover, the two extreme values of particle incidence angle were tested ($90\,^{\circ}$ and $84.29\,^{\circ}$). 
\begin{figure}[h!]
\centering
\subfigure[]{
\includegraphics[scale=0.86]{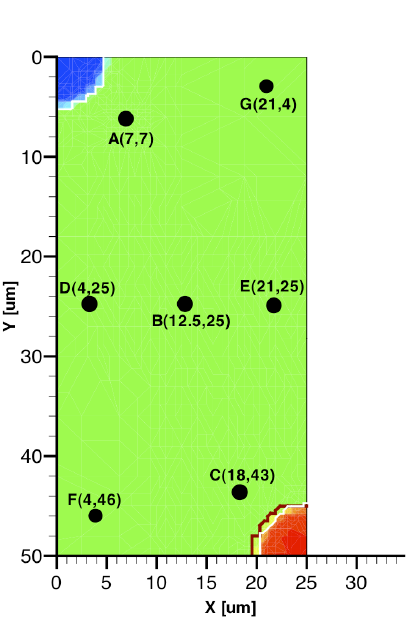}
\label{fig:F8a}
}
\subfigure[]{
\includegraphics[scale=0.86]{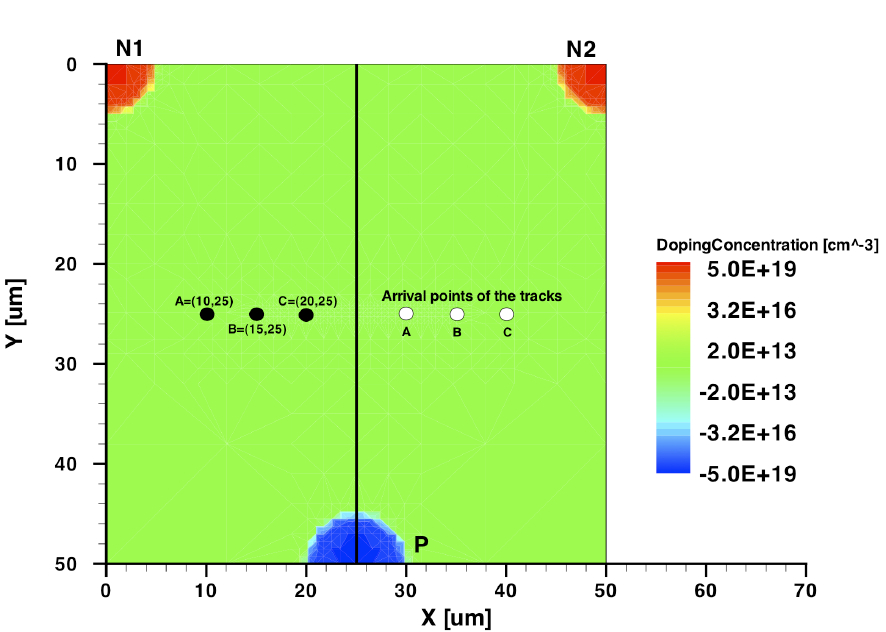}
\label{fig:F8b}
}
\caption{Front view of 4E simulated structures:  \subref{fig:F8a} structure for cluster size 1 simulations with different particle hit points, and \subref{fig:F6b} structure for cluster size 2 simulations with different particle hit and arrival points (track angle $84.29\,^{\circ}$). Similar simulation domains choice apply to the 2E and 3E devices.}
\end{figure}
To this purpose, since the probability of charge sharing is strictly related to the inclination of the particle, two different structures were simulated see Fig.\,8, which refers to the case of the 4E devices): (i) a single cell including one n$^{+}$ column and one p$^{+}$ column, for a $90\,^{\circ}$ hit angle (Fig.\,\ref{fig:F8a} and (ii) two adjacent cells including two n$^{+}$ electrodes (related to adjacent pixels) and one p$^{+}$ column, for a $84.29\,^{\circ}$ hit angle (Fig.\,\ref{fig:F8b}). Moreover, to have a better understanding of the behavior of the detectors, different particle hit points were tested and the results were combined as explained below.\\
A particle hit was simulated using the Heavy-Ion model available in the simulator: the released charge was 80 electron-hole pairs per micron and the spatial distribution was Gaussian in a region of one micrometer diameter around the track. The radiation damage was modeled using the ÒPerugiaÓ trap model \cite{Petasecca} modified as described in \cite{Pennicard}.\\
Transient simulations were performed from 0 to 100\,ns; the high leakage current was subtracted from the output current and a numerical integration was performed in order to extract the total collected charge for each simulation. In order to decide if an event was good or not, the collected charge after 20\,ns was observed and, if the value was above the threshold (3200\,e), the hit was considered valid and the value at which the integral saturated was taken as the total charge collected in that event. In the charge sharing simulations, the same procedure was adopted for both n$^{+}$ columns. \\
Since measurements showed different distributions for cluster size 1 and cluster size 2 events, simulations were combined using the following equation:
\begin{equation}
Q_{all-cluster, sim} = CS1_{\%} \times Q_{CS1} +  CS2_{\%} \times Q_{CS2}
\end{equation}
where $Q_{CS1}$ is the total collected charge for cluster size 1 events, $Q_{CS2}$ is the total collected charge for cluster size 2 events (the sum of the charge collected on two adjacent pixels) and $CS1_{\%}$ and $CS2_{\%}$ are the percentages of cluster size 1 and cluster size 2 events obtained from the measurements. In particular the 2E and 3E devices showed 90\% of cluster size 1 events and 10\% of cluster size 2 events while these percentages became 80\% and 20\% for the 4E device. Simulation results were compared with measured Most Probable Values and a very good agreement was found (see Fig.\,\ref{fig:F9}), thus confirming that the most important physical mechanisms are properly modeled in the simulations.
In order to gain a better insight into the charge collection process, several cuts were extracted from the simulations results showing different electrical quantities, among them the electric field, Shockley-Read-Hall (SRH) recombination, electron density, and hole density for different bias voltages and different depths inside the device. Let's consider as an example the 4E device. Two voltages (20\,V and 60\,V) were chosen as representative of low-bias and high-bias conditions, and the cuts were extracted from two separate regions, one in the part of the sensor where columnar electrodes overlap (z\,=\,50\,$\mu$m) and the other in the non-overlap region (z\,=\,150$\mu$m). 
This was done as these two regions of the device behave completely differently, especially after irradiation.\\
The above-mentioned quantities, relevant to a time instant 0.5\,ns after the particle hit the device perpendicularly to the wafer surface (point E in Fig.\,\ref{fig:F8a}), are shown in Figs.\,\ref{fig:F10}\,- \,\ref{fig:F12}. \\
Looking at Fig.\,\ref{fig:F10}, it is clear that for both bias voltages the electric field is strong only in the overlap region (with a much higher absolute value for 60\,V bias), whereas it is very weak in the bottom part of the sensor.\\
From Fig.\,\ref{fig:F11}, it is possible to confirm that SRH recombination is negligible in the overlap region, whereas it significantly affects those carriers generated by the impinging particle in the non-overlap region.\\
\clearpage
\newpage
\begin{figure*}[h!]
\centering
\includegraphics[scale=0.3]{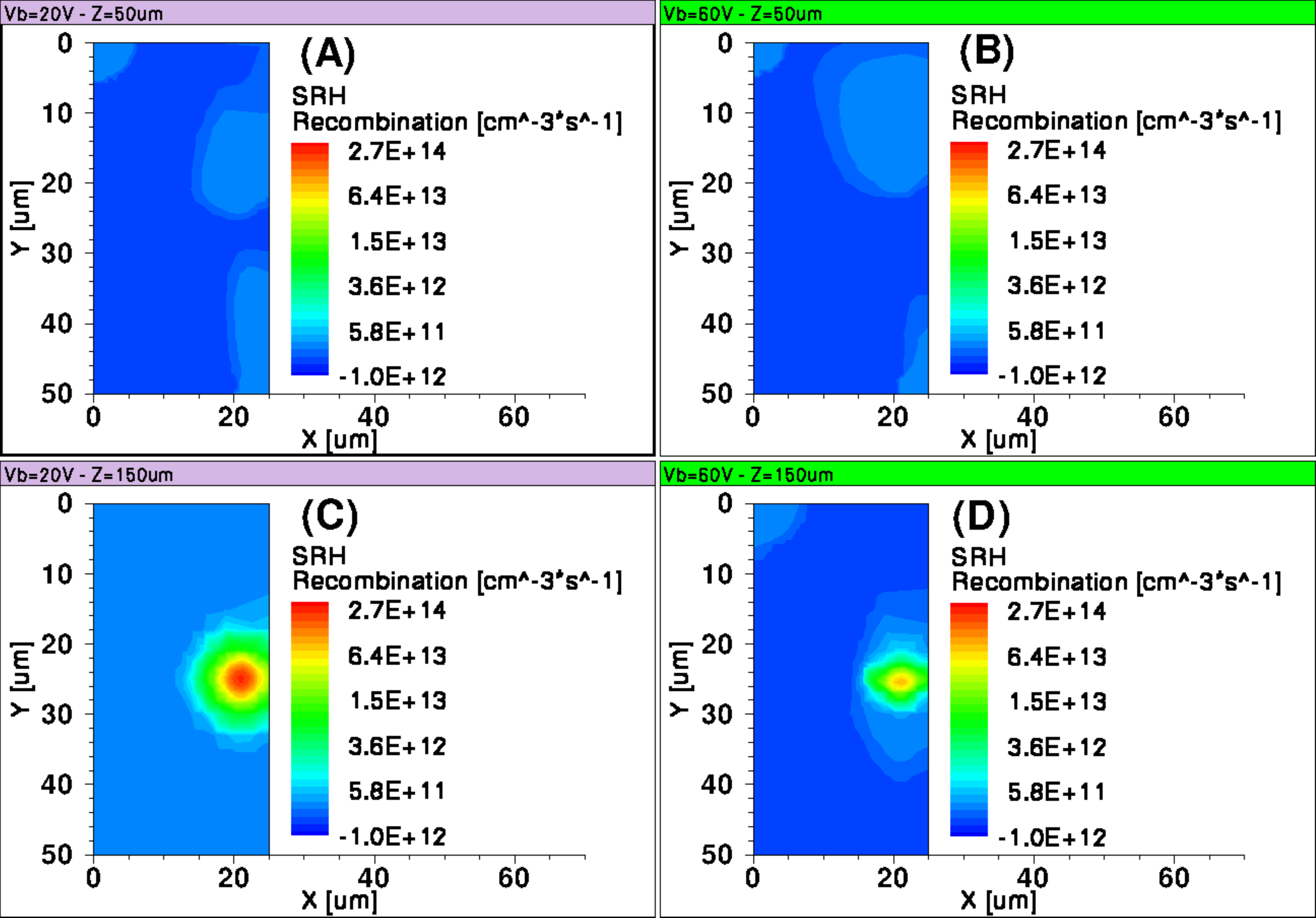}
\caption{2d cuts in the 4E sensor showing the SRH recombination at two different depths (Z) and two bias voltages (V$_{\mathrm{b}}$): (A) Z\,=\,50\,$\mu$m, V$_{\mathrm{b}}$\,=\,20\,V;  (B) Z\,=\,50\,$\mu$m, V$_{\mathrm{b}}$\,=\,60\,V;  (C) Z\,=\,150\,$\mu$m, V$_{\mathrm{b}}$\,=\,20\,V; (D) Z\,=\,150\,$\mu$m, V$_{\mathrm{b}}$\,=\,60\,V. Data are relevant to a time instant 0.5\,ns after a particle hit the device perpendicularly to the wafer surface (point E in \ref{fig:F8a}).}
\label{fig:F11}
\end{figure*}
\clearpage
\newpage
\begin{figure*}[th!]
\centering
\includegraphics[scale=0.3]{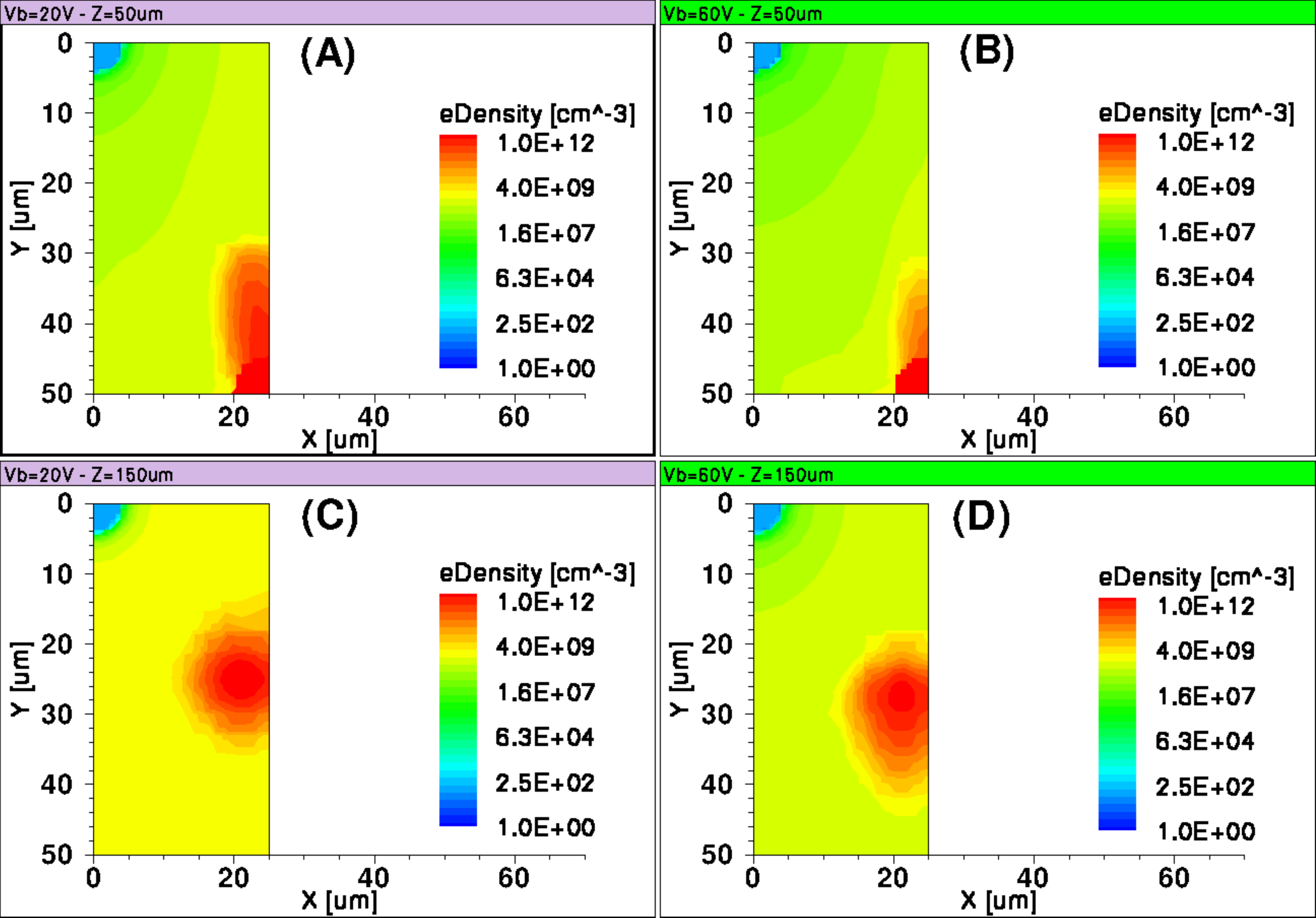}
\caption{2d cuts in the 4E sensor showing the electron density at two different depths (Z) and two bias voltages (V$_{\mathrm{b}}$): (A) Z\,=\,50\,$\mu$m, V$_{\mathrm{b}}$\,=\,20\,V;  (B) Z\,=\,50\,$\mu$m, V$_{\mathrm{b}}$\,=\,60\,V;  (C) Z\,=\,150\,$\mu$m, V$_{\mathrm{b}}$\,=\,20\,V; (D) Z\,=\,150\,$\mu$m, V$_{\mathrm{b}}$\,=\,60\,V. Data are relevant to a time instant 0.5\,ns after a particle hit the device perpendicularly to the wafer surface (point E in \ref{fig:F8a}).}
\label{fig:F12}
\end{figure*}
%
%\begin{figure*}[h!]
%\centering
%\includegraphics[scale=0.3]{F13.pdf}
%\caption{2d cuts showing the hole density at two different depths (Z) and two bias voltages (V$_{\mathrm{b}}$): (A) Z\,=\,50\,$\mu$m, V$_{\mathrm{b}}$\,=\,20\,V;  (B) Z\,=\,50\,$\mu$m, V$_{\mathrm{b}}$\,=\,60\,V;  (C) Z\,=\,150\,$\mu$m, V$_{\mathrm{b}}$\,=\,20\,V; (D) Z\,=\,150\,$\mu$m, V$_{\mathrm{b}}$\,=\,60\,V. Data are relevant to a time instant 0.5\,ns after a particle hit the device perpendicularly to the wafer surface (point E in \ref{fig:F8a}).}
%\label{fig:F13}
%\end{figure*}
%
%
%\clearpage
%\newpage
%
\begin{figure*}[t]
\centering
\subfigure[]{
\includegraphics[scale=0.7]{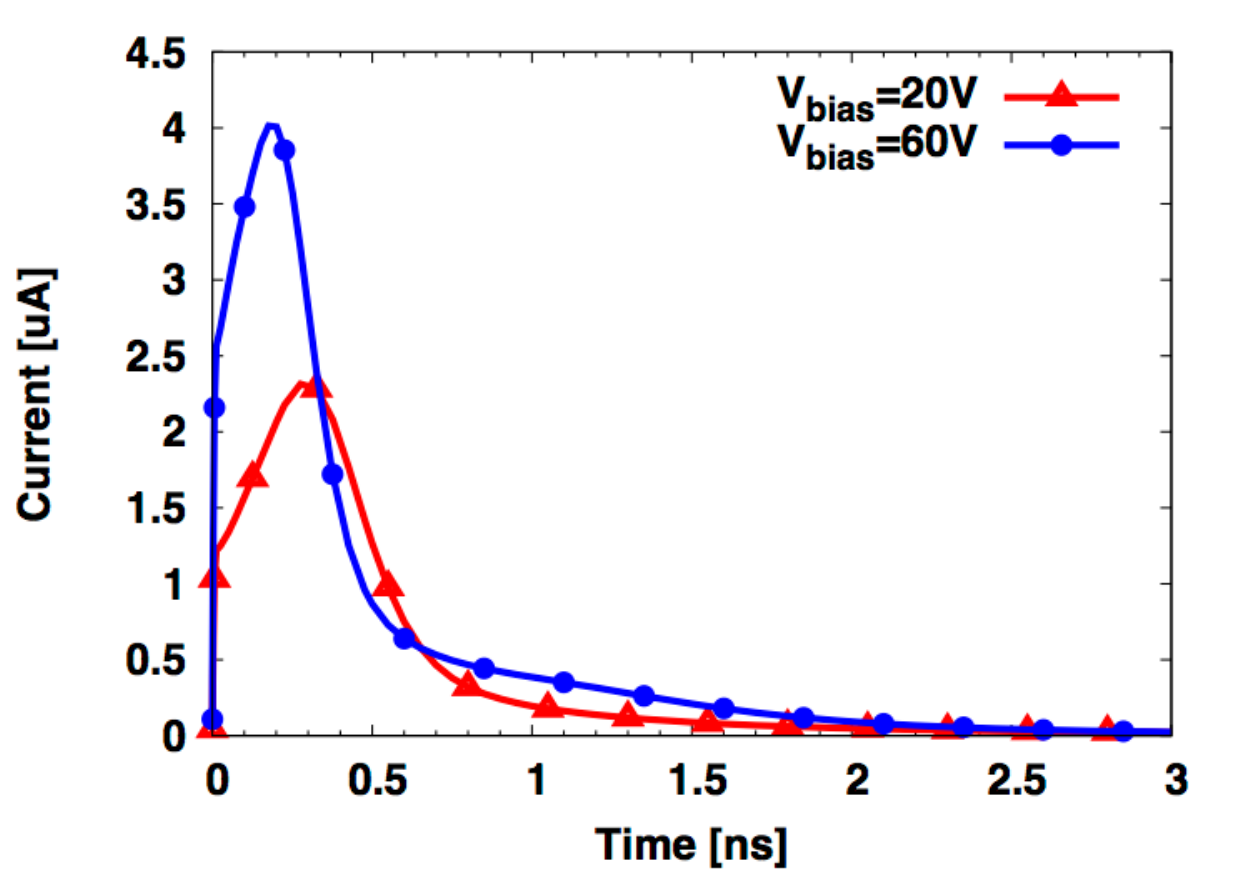}
\label{fig:F14a}
}
\subfigure[]{
\includegraphics[scale=0.7]{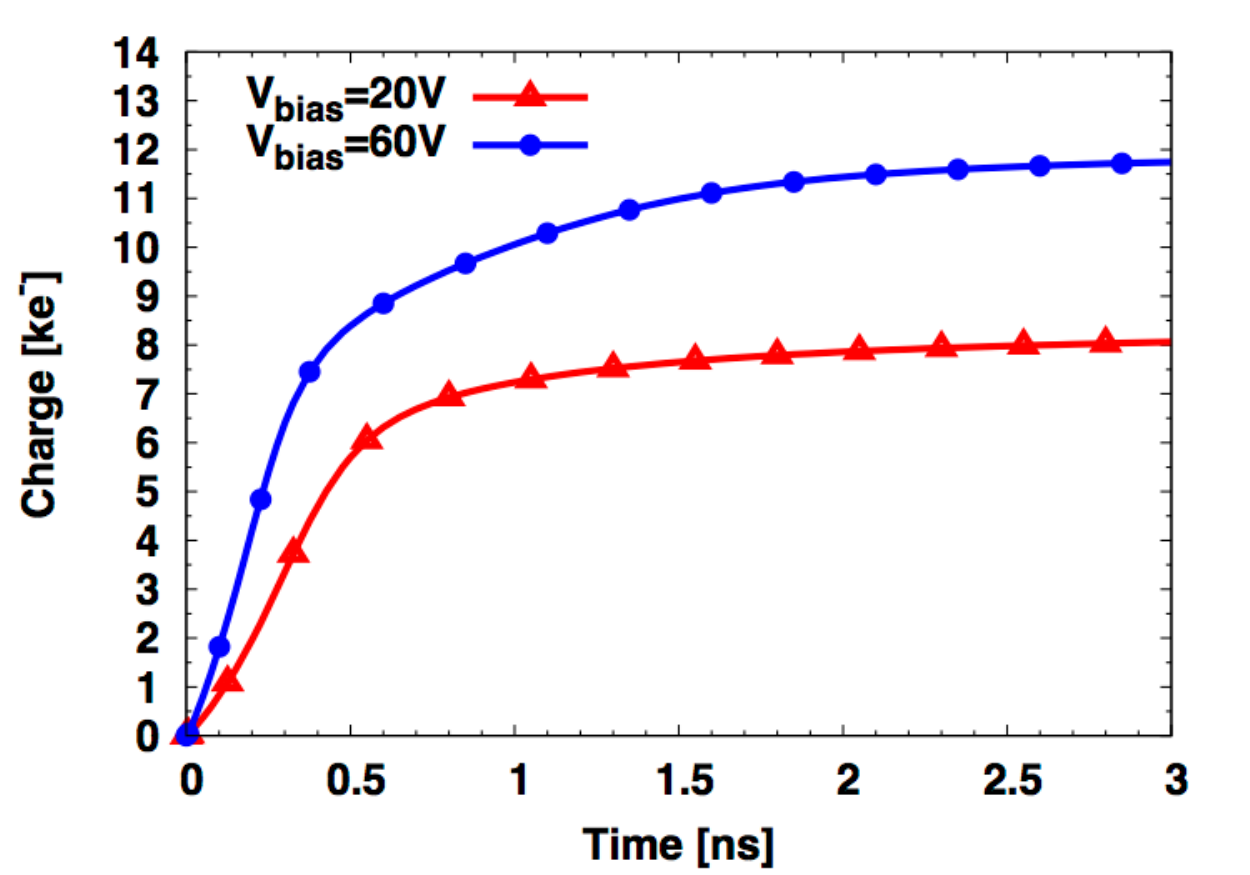}
\label{fig:F14b}
}
\caption{Simulated output currents in the 4E sensor at 20\,V and 60\,V bias \subref{fig:F14a} and corresponding time integrals \subref{fig:F14b}.}
\end{figure*}
\clearpage
\newpage
Both the distributions of the electric field distribution and the SRH recombination can be explained considering the way the depletion region extends inside this type of devices: full depletion occurs at relatively low bias voltages in the overlap region, whereas much larger voltages would be required to fully deplete the bottom part of the device. \\
Given that part of the charge generated by the impinging particle is lost because of recombination, it is important to understand how the rest of the carriers move inside the device and how they contribute to the signal formation. Fig.\,\ref{fig:F12} shows the electrons density distributions for the biasing conditions, depths and time instant.
From the electron density point of view, it is clear that carriers generated in the top part of the detector are collected almost immediately owing to the strong electric field. The effect of the bias voltage is clearly visible because the electrons are collected faster at 60\,V than at 20\,V in this region. The situation in the bottom part of the device is different: the electric field is much weaker and carriers must drift (or even diffuse in non depleted regions) both laterally and vertically toward the tip of the n$^{+}$ column. Since distances are longer and the field is weaker, carriers in this region will require more time to be collected, drastically increasing the probability of being trapped or recombining. The minor changes in the bottom part of the device between 20\,V and 60\,V indicate that the device has not reached an optimal working point. The previous considerations can also be applied to what happens for the hole density (not shown). Again, holes are collected much faster in the overlap region than in the bottom region.\\
The above analysis is in good agreement with simulation results in terms of transient output currents (Fig.\,\ref{fig:F14a}) and their time integrals (Fig.\,\ref{fig:F14b}) for the two considered biases (20\,V and 60\,V). As already mentioned, for higher bias voltages, electrons and holes in the overlap region suffer from less trapping and this translates into a pulse with higher amplitude and faster evolution thanks to the higher electric field (both electrons and holes generated in the overlap region concur in the formation of the fast peak). 
Apart from the difference in the peak amplitude and time, a change of the shape of the pulse tails can be observed: 
at 60\,V of bias, the pulse tail extends to a slightly longer time, a fact that is also confirmed by the delayed saturation of the corresponding time integral curve in Fig.\,\ref{fig:F14b}. The change of the shape of the pulse tails is related to the fact that at 60\,V of bias, the bottom part of the detector suffers from less recombination and the charge generated there also contributes to the output signal. Since the field in this region is weaker than in the top, the drift of the carriers will be slower, hence the presence of the long tail.
Similar considerations apply to the other types of sensors, that are not reported here for the sake of conciseness.

%% file: conclusion.tex
\section{Conclusion}
\label{sec:conclusion}

We have reported on selected results from the functional characterization of irradiated 3D-DDTC pixel sensors fabricated at FBK. Sensors with different pixel configurations have been assembled with the ATLAS FEI3 read-out chip and irradiated with protons up to very large fluences. Experimental results from measurements carried out in the laboratory have been discussed and compared to those obtained before irradiation. As expected, the breakdown voltage was generally increased by a few tens of Volts with respect to the pre-irradiation values, allowing to operate the sensors at a bias voltage high enough to achieve at least lateral depletion between the columns. The peak of the collected charge in response to $\gamma$-rays from an $^{241}$Am source is not significantly degraded with respect to the values obtained before irradiation, provided that the bias voltage is high enough to effectively counteract trapping in the top region of the devices where columns overlap. On the contrary, the most probable value of the collected charge for tests with $\beta$-particles from a $^{90}$Sr source is found to be more sensitive to trapping, due to the lower charge collection efficiency from the bottom region of the devices, where columns do not overlap, in good agreement with TCAD simulations. Nevertheless, the signal efficiency, defined as the ratio of the MPV of the collected charge after irradiation and before irradiation, remains at acceptable levels: for the best sample irradiated at 1\,x\,10$^{15}$\,n$_{\mathrm{eq}}$cm$^{-2}$  it is  about 76\,\% at 100\,V, and for the best sample irradiated at 2\,x\,10$^{15}$\,n$_{\mathrm{eq}}$cm$^{-2}$ it is about 64\,\% at 120\,V. These results are very encouraging since the tested samples have a rather short column overlap, less than one half of the sensor thickness. Hence, there is still wide room for performance improvement by etching deeper junction column depths. New 3D-DDTC sensors fabricated at FBK have indeed passing-through columns and are expected to be very radiation hard. 